\documentclass[conference]{IEEEtran}
\usepackage{overpic}
\usepackage{epsfig}
\usepackage{amssymb,amsmath}
\usepackage{fancybox}
\usepackage{pstricks}
\usepackage{graphicx}

\def\defineTColor#1#2{
  \newpsstyle{#1}{%
    fillstyle=vlines,hatchcolor=#2,%
    hatchwidth=0.1\pslinewidth,hatchsep=1\pslinewidth%
}}
\defineTColor{tCyan}{cyan}
\newcommand{\CK}{\v Cerenkov}
\newcommand{\myframebox}[3]{%
\psframebox[fillstyle=solid,fillcolor=#2]{%
\parbox{#1\textwidth}{%
       \black \begin{flushleft}{#3}\end{flushleft}
        }}}

\begin{document}

\title{Cosmic ray velocity and electric charge measurements with the AMS/RICH
                  detector: prototype results}

\author{\authorblockN{Lu\'{\i}sa Arruda, Fernando Bar\~ao, Patr\'{\i}cia Gon\c calves, Rui Pereira}
\authorblockA{LIP/IST \\
         Av. Elias Garcia, 14, 1$^o$ andar\\
         1000-149 Lisboa, Portugal \\
         e-mail: luisa@lip.pt}}

\maketitle

\begin{abstract}
The Alpha Magnetic Spectrometer (AMS) to be installed on the International
Space Station (ISS) will measure charged cosmic ray spectra of elements up to
iron, in the rigidity range from 1\,GV to 1\,TV, for at least three years.
AMS is a large angular spectrometer composed of different subdetectors,
including a proximity focusing Ring Imaging CHerenkov (RICH) detector. This
will be equipped with a mixed radiator made of
aerogel and sodium fluoride (NaF), a lateral conical
mirror and a detection plane made of 680 photomultipliers coupled to light
guides. The RICH detector allows measurements of particle's electric charge up to iron, and particle's velocity. Two possible methods for reconstructing the
\CK\ angle and the electric charge with the RICH will be discussed. 

A RICH prototype consisting of a detection matrix with 96 photomultipliers, a segment of a conical mirror and samples of the radiator materials
was built and its performance was evaluated using ion beam data. Results from
the last test beam performed with ion fragments resulting from the collision
of a 158\,GeV/c/nucleon primary beam of indium ions (CERN SPS) on a lead
target are reported. The large amount of collected data allowed to test and
characterize different aerogel samples and the NaF radiator. In
addition, the reflectivity of the mirror was evaluated. The data analysis confirms the design goals.
\end{abstract}

\section{The AMS-02 experiment}
\label{AMS-02}
\begin{figure}[htb]
\begin{center}
\vspace{-0.45cm}
 \epsfig{file=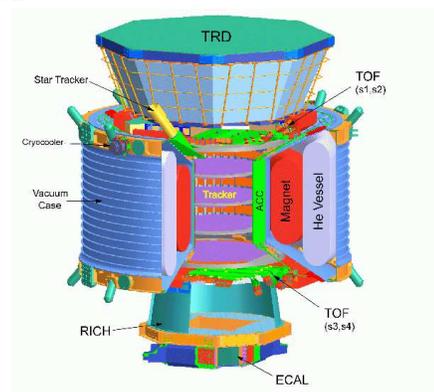,width=0.65\linewidth,clip=,bbllx=0,bblly=0,bburx=612,bbury=550} 
\caption{A whole view of the AMS Spectrometer.}
\end{center}  
\label{fig:ams}
\end{figure}

The Alpha Magnetic Spectrometer \cite{bib:ams02-note} (AMS) is a particle detector
to be installed in the International Space Station (ISS) for at least three years.
The spectrometer  will be able to measure
the rigidity ($R\equiv pc/ |Z| e$), the charge ($Z$),
the velocity ($\beta$) and the energy ($E$) of cosmic rays from some MeV up
to $\sim$1\,TeV within a geometrical acceptance of $ \sim$0.5\,m$^2$.sr.
\mbox{Figure \ref{fig:ams}} shows a schematic view of the AMS spectrometer.
At both ends of the AMS spectrometer exist the 
Transition Radiation Detector
(TRD) (top) and the Electromagnetic Calorimeter (ECAL) (bottom). Both will
provide AMS with capability to discriminate between leptons and hadrons. 
Additionally the calorimeter will trigger and detect photons.
The TRD will be followed by the first of the four Time-of-Flight (TOF) system  scintillator planes. 
The TOF \mbox{system \cite{bib:ams02-tof}} is composed of four roughly circular planes of 12\,cm wide
scintillator paddles, one pair of planes above the magnet, the upper TOF,
and one pair bellow, the lower TOF. There will be a total of 34 paddles. The
TOF will provide a fast trigger within 200\,ns, charge and velocity measurements for charged particles, 
as well as information on their direction of incidence. The TOF operation at
regions with very intense magnetic fields forces the use of shielded fine mesh
phototubes and the optimization of the light guides geometry, with some of
them twisted and bent. Moreover the system guarantees redundancy, with two photomultipliers
on each end of the paddles and double redundant electronics. A time
resolution of 140\,ps for protons is \mbox{expected \cite{bib:ams02-tof}}.

The tracking system will be surrounded by veto counters and
embedded in a magnetic field of about 0.9\,Tesla produced by a 
superconducting magnet.
It will consist on a Silicon \mbox{Tracker \cite{bib:ams02-tracker}}, 
made of 8 layers of double sided silicon 
sensors with a total area of $\sim$6.7\,m$^2$. There will be a total of
$\sim$2500 silicon sensors arranged on 192 ladders. The position of the charged particles crossing
the tracker layers is measured with a precision of $\sim$10\,$\mu$m along
the bending plane and $\sim$30\,$\mu$m
on the transverse direction. 
With a bending power (BL$^2$) of around 0.9\,T.m$^2$, particles rigidity is measured    
with an accuracy better than 2\% up to 20\,GV and the maximal detectable rigidity is
around 1-2\,TV. Electric charge is also measured from energy deposition up
to Z$\sim$26.

The Ring Imaging \CK\ Detector (RICH) \cite{bib:baraoICRC}
will be located right after the last TOF plane and before the electromagnetic
calorimeter. It will be described in detail in next section.

 The long stay of AMS in space and its large acceptance will allow the accumulation of a large
statistic of events increasing in several orders of magnitude the sensitivity
of the proposed physical measurements. With an average collection rate of
1000 events per second, a total of $10^9$ protons per year
and around $10^4$ antiprotons will be accumulated.

The main goals of the AMS-02 experiment are:

\begin{itemize}

\item A precise measurement of the charged cosmic ray spectrum between
\mbox{$\sim$100\,MeV} and $\sim$1\,TeV, and the detection of photons up to
a few hundred GeV;

\item A search for heavy antinuclei ($Z \ge 2$), which if discovered would
signal the existence of primordial antimatter;

\item Search for non-baryonic dark matter through the detection of annihilation products appearing as anomalies of the cosmic-ray spectra (e$^+$, $\bar{p}$, $\gamma$ and $\bar{d}$);

\end{itemize}

\section{The AMS RICH detector}
The RICH is a proximity focusing device with a dual radiator configuration on
the top made of 92 aerogel 25\,mm thick tiles with a refractive index 1.050 and 
sodium fluoride (NaF) tiles with a  thickness of 5\,mm in the center
covering an area of 34$\times$34\,cm$^2$.  
The NaF placement prevents the loss of photons in the hole existing in the center of the readout plane  
($64\times64$ cm$^2$), in front of the ECAL calorimeter located below. The
radiator tiles are supported by a 1\,mm thick layer of methacrylate (n=1.5)
free of UV absorbing additives.

The detection matrix is composed of 680 multiplixelized photon readout cells
each consisting of a photomultiplier coupled to a light guide, HV divider plus
front-end (FE) electronics, all housed and potted in a plastic shell and then
enclosed in a magnetic shielding with a thickness varying from 0.8 to
1.2\,mm. 
The photon detection is made with an array of multianode Hamamatsu tubes (R7600-00-M16) with a spectral 
response ranging from 300 to 650\,nm and a maximum quantum efficiency at $\lambda$$\sim$420\,nm.
To increase the photon collection efficiency, a light guide consisting of 16 solid acrylic pipes 
glued to a thin top layer (1\,mm) was produced. 
It is optically coupled to the active area of phototube cathode through a $1$ mm flexible optical pad.
With a total height of $31$ mm and a collecting surface of $34\times34$ mm$^2$, it presents a readout pixel size 
of $8.5$ mm and a pitch of 37\,mm.  
The light guide is mechanically attached through nylon wires to the
photomultiplier polycarbonate housing.

A high reflectivity conical mirror
surrounds the whole set. The mirror was included to increase the device
acceptance since around 33\% of the
aerogel generated photons impact on the mirror. It consists of a carbon fiber
reinforced composite substrate with a multilayer coating made of aluminium
and SiO$_2$ deposited on the inner surface. This ensures a reflectivity
higher than 85\% for 420\,nm wavelength photons.

\begin{figure}[htb]

\center

\vspace{-0.3cm}

\mbox{\epsfig{file=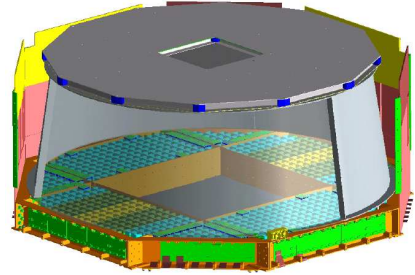,width=0.3\textwidth,clip=}}


\caption{Schematic view of the RICH detector.\label{fig:rich}}

\end{figure}

The RICH has a truncated conical shape with an
expansion height of 46.3\,cm, a top radius of 60\,cm and a bottom radius of
67\,cm. The total height of the detector is 60.5\,cm and it covers 80\% of
the AMS magnet acceptance. \mbox{Figure \ref{fig:rich}} shows a schematic view of the RICH detector. 

RICH was designed to measure the velocity
($\beta\equiv v/c$) of singly charged particles with a resolution $\Delta\beta/\beta$ of 0.1$\%$,
to extend the charge separation (Z) up to iron (Z=26), to contribute to $e/p$ separation 
and to albedo rejection.

\begin{figure}[htb]

\center
\begin{tabular}{cc}

\epsfig{file=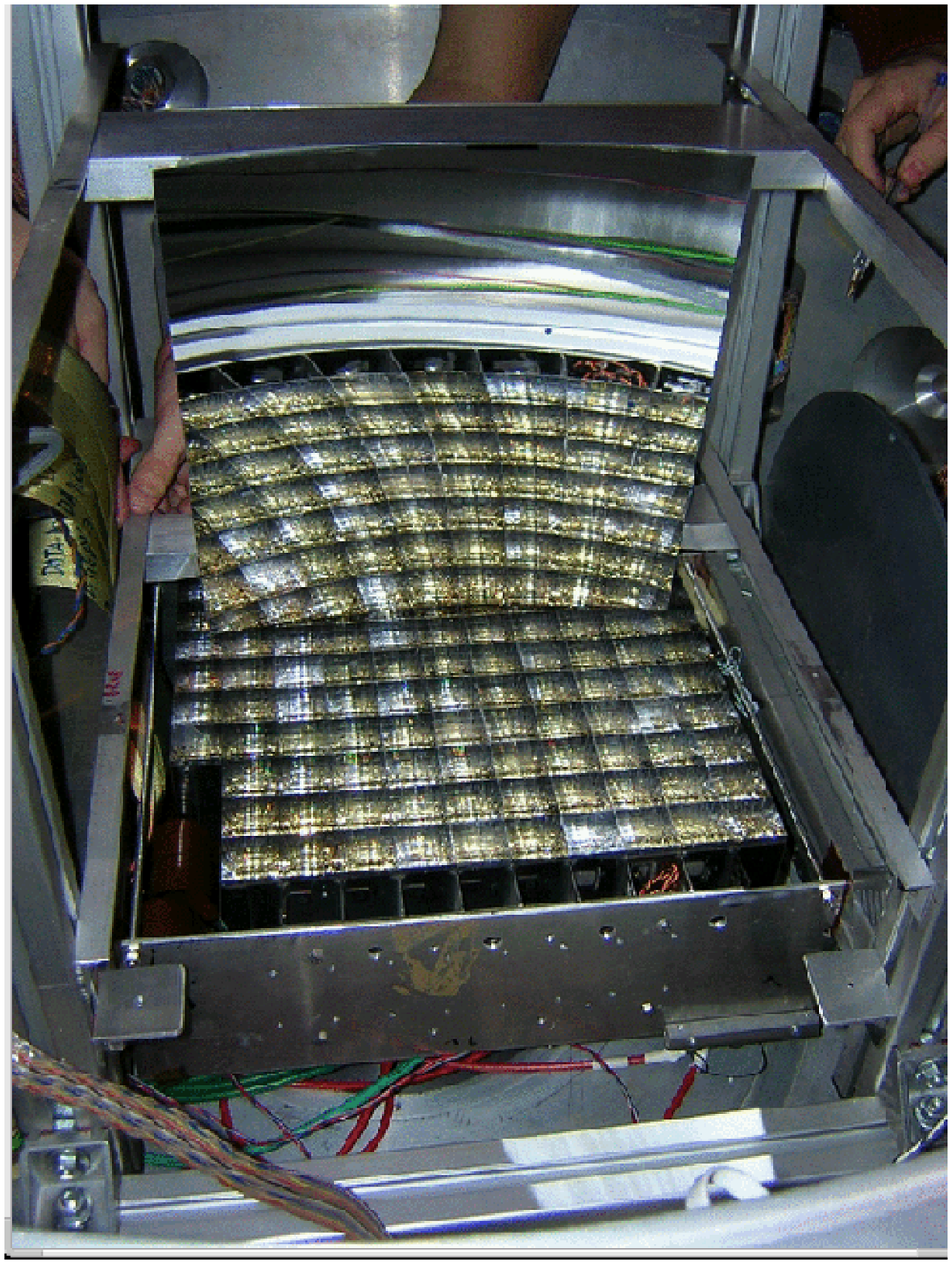,width=0.23\textwidth,clip=}
&
\epsfig{file=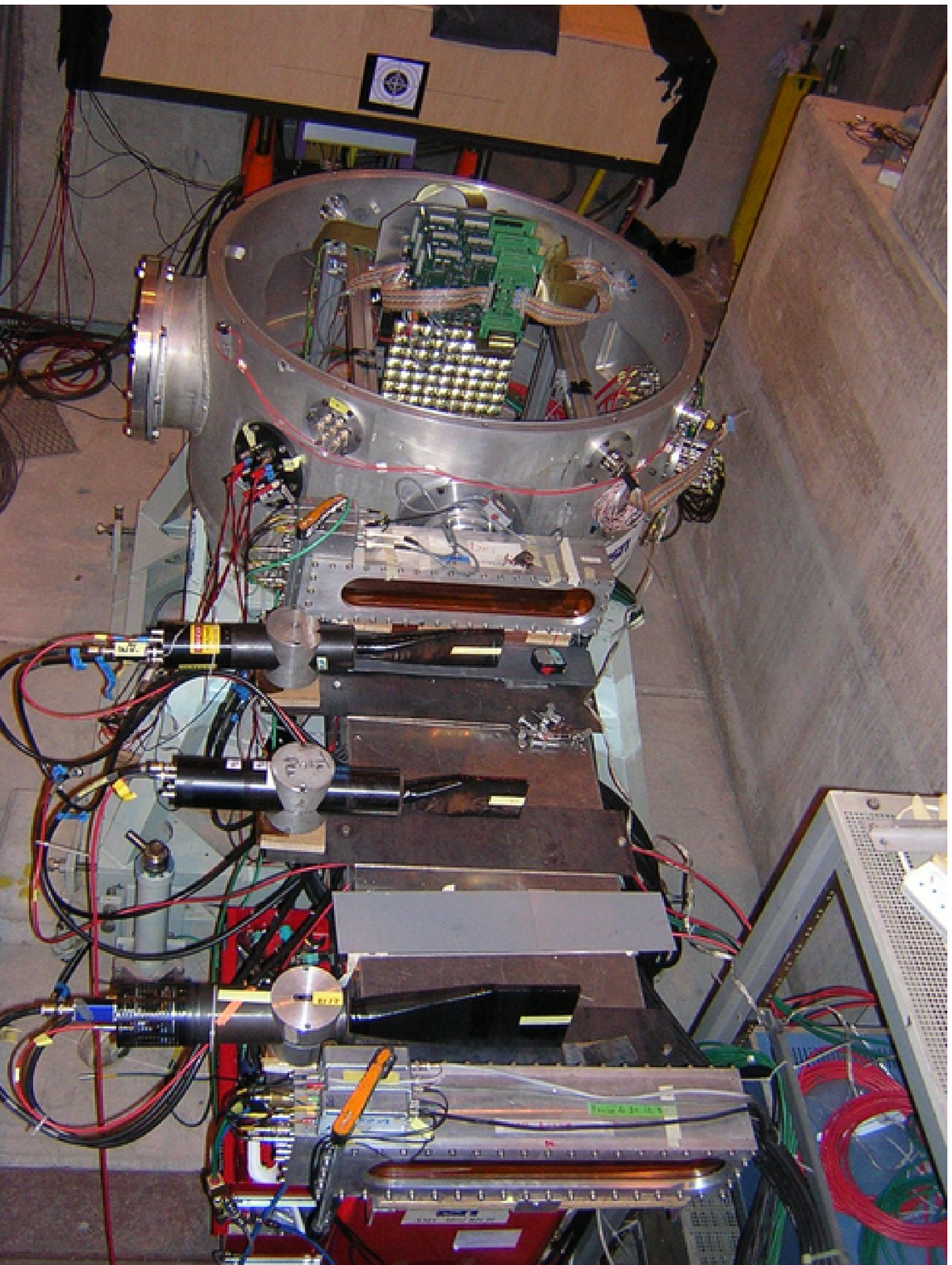,width=0.23\textwidth,clip=}
\vspace{-0.15cm}
\end{tabular}   
\caption{Protype with reflector (left). Top view of the test beam 2003
  experimental setup using CERN SPS facility (right).\label{fig:tb03}}

\end{figure}

In order to validate the RICH design, a prototype with an array of
9$\times$11 cells filled with 96 photomultiplier readout units similar to part of the matrix of the final model was constructed. The
performance of this prototype has been tested with cosmic muons and with a
beam of secondary ions at the CERN SPS produced by fragmentation of a primary
beam in 2002 and 2003. The light guides used were prototypes with a slightly
smaller collecting area (31$\times$31\,mm$^2$).
 Different samples of the radiator materials were tested and placed at an
 adjustable supporting structure. Different expansion heights were set
 in order to have fully contained photon rings on the detection matrix like
 in the flight design.
A segment of a conical mirror with 1/12 of the final azimuthal coverage, which is shown in left picture of \mbox{Figure
  \ref{fig:tb03}}, was also tested. 

The RICH assembly has already started at CIEMAT in Spain and is foreseen to
be finished in July 2007. A rectangular grid has already been assembled and has been
subject to a mechanical fit test, functional tests, vibration tests and
vacuum tests. The other grids will follow. The refractive index of the
aerogel tiles is being measured and the radiator container was subjected to a
mechanical test. The final integration of RICH in AMS will take place at CERN
in 2008.

\section{Beta ($\beta$) and charge ($Z$) reconstruction}
\label{BETAZ}
A charged particle crossing a dielectric material of refractive index $n$
with a velocity $\beta$, greater than the speed of light in that 
medium, emits photons.
The aperture angle of the emitted photons with respect to the 
radiating particle is known as the \CK\ angle, $\theta_c$,
and it is given by (see Ref.~\cite{bib:NIM})
\vspace{-0.1cm}
\begin{equation}
\cos\theta_c = \frac{1}{\beta~n}
\label{eq1}
\end{equation}

\begin{figure}
\begin{center}
\vspace{-0.8cm}
\begin{tabular}{cc}
\scalebox{0.27}{\includegraphics{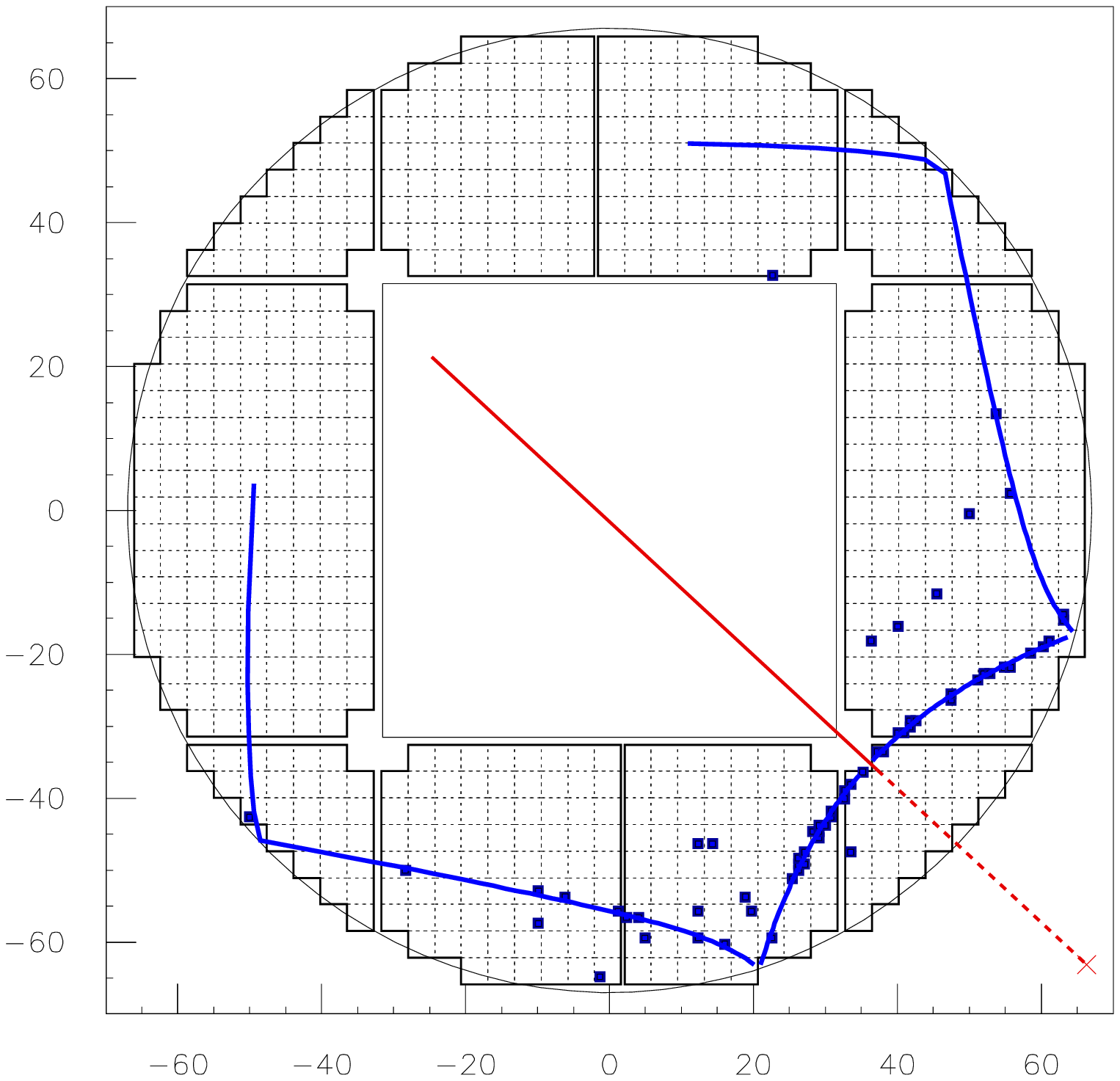}}
&
\hspace{-1.0cm}
\scalebox{0.28}{\includegraphics[bb=0 0 482 482]{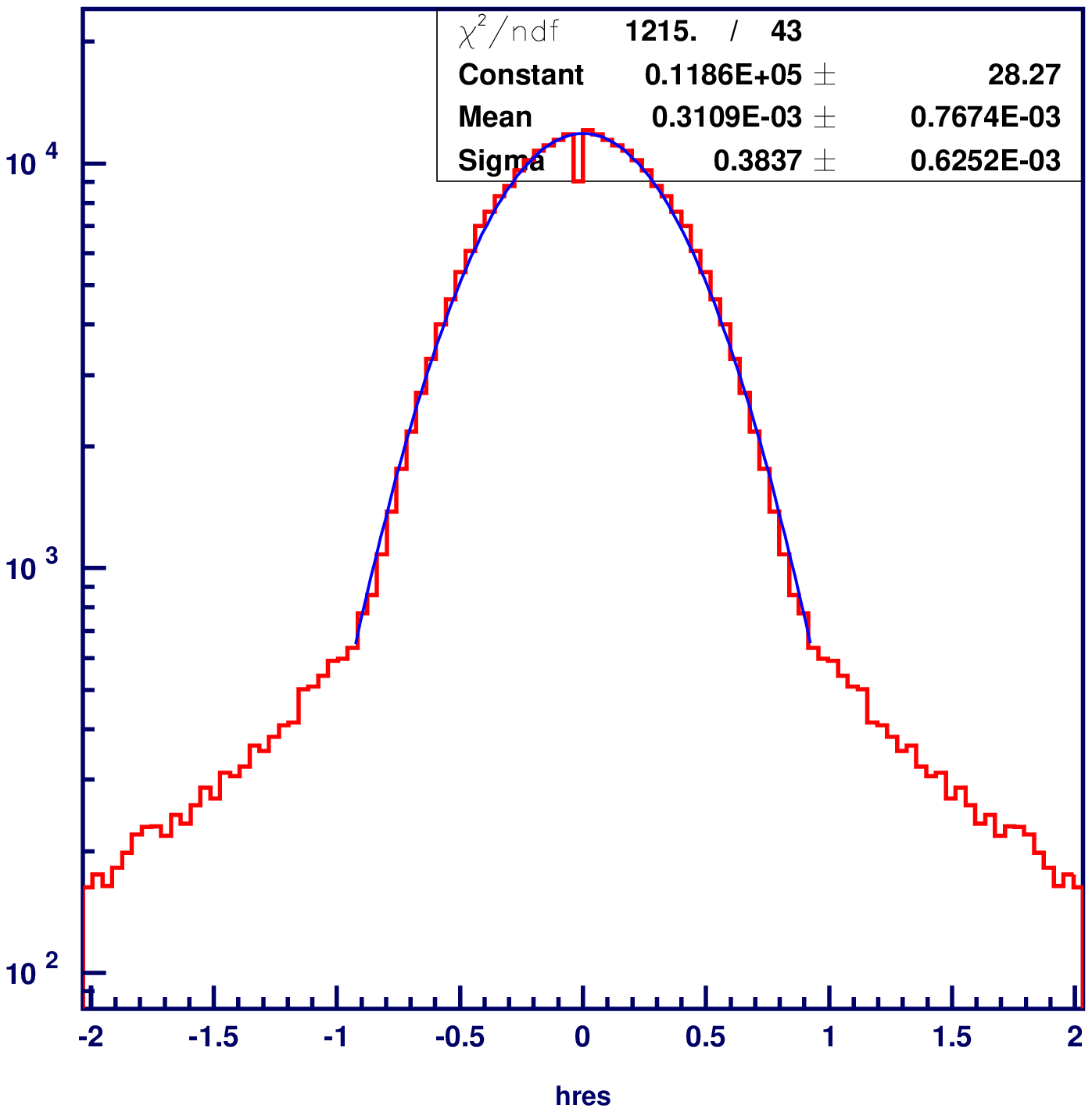}}
\end{tabular}  
\caption{Beryllium event display generated in a NaF radiator. The reconstructed photon pattern (full line)
  includes both reflected and non-reflected branches. The outer circular
  line corresponds to the lower 
boundary of the conical mirror. The square is the limit of the non-active
region (left). Hit's residuals distribution belonging to photon rings
  generated in an aerogel radiator, n=1.05, 2.5\,cm thick. \label{fig:disp}}
\end{center}
\end{figure}

It follows that the velocity of the particle, $\beta$, is straightforward
derived from the \CK\ angle reconstruction, which is based on a fit to the
pattern of the detected photons.
Complex photon patterns can occur at the detector plane 
due to mirror reflected photons, as can be seen on the left display of \mbox{Figure
\ref{fig:disp}}. The event shown is generated by a simulated beryllium
nucleus crossing the NaF radiator. 
The \CK\ angle reconstruction procedure 
relies on the higly accurate information (see \mbox{Section \ref{AMS-02}}) of the particle direction 
($\theta$,$\phi$) provided by the tracker.
The tagging of the hits signaling the passage 
of the particle through the solid light guides in the detection plane 
provides an additional track ele\-ment, however, those hits are excluded from the reconstruction.
The best value of $\theta_c$ will result from the maximization of a Likelihood function, 
built as the product of the probabilities, $p_i$, that the detected hits belong to a given (hypothetical)
\CK\ photon pattern ring,
\vspace{-0.2cm}
\begin{equation}
L(\theta_c) = \prod_{i=1}^{nhits} p_i^{s_i}  \left[ r_i(\theta_c) \right].
\label{eq:likelihood}
\vspace{-0.3cm}
\end{equation}

This probability takes into account $r_i$, the closest distance of the hit to
the \CK\ pattern and $s_i$ the signal strength. 
The probability of a hit belonging to the pattern is obtained by taking into
account that it can either be part of the noise (essentially flat) or of the
\CK\ pattern
(Gaussian distributed).  Expressing $b$ as the photon background fraction,
$D$ as the detector's dimensions and $\sigma$ as the width of the residuals
distribution (as can be seen in right plot of \mbox{Figure \ref{fig:disp}}), we can write:
\begin{equation}
P_i = (1-b)G(\sigma ;r_i) + \frac{b}{D}
\end{equation}
%
where $b=0.5122$ and $0.105$ respectively for
aerogel and NaF in the prototype setup. The detector's dimension is defined as D=100\,cm for
flight setup and 20\,cm for prototype setup.
For a more complete description of the method see Ref.~\cite{bib:NIM}.

The \CK\ photons produced in the radiator are uniformly emitted along the particle path 
inside the dielectric medium, $L$, and their number per unit of energy 
depends on the particle's charge, $Z$, and velocity, $\beta$, and on 
the refractive index, $n$, according to the expression:\\
\begin{equation}
\frac{dN_{\gamma}}{dE} \propto Z^2 L \left( 1-\frac{1}{\beta^2 n^2} \right)
\label{eq:dnde}
\end{equation}

Therefore to reconstruct the charge the following procedure is required:
\begin{itemize}
\item \CK\ angle reconstruction. 
\item Estimation of the particle path, $L$, which relies on the information of the particle 
direction provided by the tracker.
\item Counting the number of photoelectrons.
The number of photoelectrons related to the \CK\ ring has to be counted within a 
fiducial area, in order to exclude the uncorrelated background. In particular
it ensures the exclusion of photons which are scattered in the radiator.  
A distance of 13\,mm to the ring was defined as the limit for photoelectron
counting, corresponding to a ring width of $\sim$5 pixels.
\item Evaluation of the photon detection efficiency.
The number of radiated photons ($N_{\gamma}$) which will be detected ($n_{p.e}$) 
is reduced due to 
the interactions with the radiator ($\varepsilon_{rad}$), the photon ring acceptance 
($\varepsilon_{geo}$), light guide efficiency ($\varepsilon_{lg}$) and photomultiplier efficiency 
($\varepsilon_{pmt}$).
\begin{equation}
n_{p.e.} \sim N_{\gamma}~\varepsilon_{rad}~\varepsilon_{geo}~\varepsilon_{lg}~\varepsilon_{pmt}  
\end{equation}
\end{itemize}

The charge is then calculated according to expression \ref{eq:dnde}, 
where the normalization constant can be evaluated from a calibrated beam of
charged particles.
For a more complete description of the charge reconstruction method see Ref.~\cite{bib:NIM}.

\section{Results with the RICH prototype}

The RICH prototype was subject to cosmic muons and to in-beam tests using
secondary nuclei from fragmentation of 20\,GeV/c/nucleon lead (Pb) ions in a
beryllium target
and 158\,GeV/c/nucleon indium nuclei in a Pb target from the CERN SPS in 2002
and 2003, respectively \cite{bib:tbnim}. 

In 2003 a monocromatic particle beam with a momentum resolution
0.15\%$\le\Delta \textrm{P}/\textrm{P}\le$1.5\% was obtained. The optics of the
line was tuned to provide a beam as parallel as possible, with a divergence
less than 1\,mrad. The beam section was $\sim$1\,mm$^2$ for the narrow beam
runs and $\sim$1\,cm$^2$ for the spread beam runs.

The beam nuclear composition could be selected according to the desired A/Z
value of the fragmentation products by setting the beam line rigidity at the
appropriate value. Three main selection
values were established: A/Z=2 ($^4$He, $^6$Li, $^{10}$B,
$^{12}$C,...); A/Z=2.25 to enhance the $^9$Be peak and A/Z=2.35 to enhance
the indium peak. 

Figure \ref{fig:tb03} (right) shows a general view of the
2003 test beam setup in the experimental area H8-SPS at CERN. The prototype
was placed inside a light-tight container. 
The setup was completed with AMS silicon tracker layers placed upstream in
the beam, a TOF prototype placed downstream, two multi-wire proportional 
chambers, two organic scintillator counters and during a certain period a
plastic \CK\ counter. The two scintillators placed $\sim$1\,m apart in front of the
prototype cointainer, provided the DAQ trigger as well as an
independent charge measurement. The silicon tracker prototype provided a very
precise measurement of the particle track parameters for the event
reconstruction as well as an external selection of charge.

The purposes of the tests were testing flight front-end electronics, characterize the
performance of the aerogel and the NaF radiators, estimating the mirror
reflectivity and evaluating the global functionality of the prototype.
A total number of 11 million events were recorded during eleven days.       
Different particle incidences were obtained by rotating the prototype setup
with respect to the beam line (0$^{\textrm{\scriptsize{o}}}$, 5$^{\textrm{\scriptsize{o}}}$,
10$^{\textrm{\scriptsize{o}}}$,15$^{\textrm{\scriptsize{o}}}$,20$^{\textrm{\scriptsize{o}}}$).

The event selection was mainly intended to remove wrongly reconstructed
tracks, events with clusterized hits and events arising from
later fragmentation. First, consistency between the external determination of the
track transverse coordinates and the estimation from the reconstructed ring
is required. Then events with more than one particle cluster in the detection
matrix are
rejected. Futhermore, the Kolmogorov probability of the event is calculated
requiring an uniform azimuthal distribution of the ring hits.
 
The evaluation of the aerogel samples in order to make a final radiator choice 
was one of the key issues of these tests.
Different production batches from two manufacturers, Matsushita Electric Co. (MEC) and   
Catalysis Institute of Novosibirsk (CIN) with different refractive indexes,
1.03 and 1.05, were analyzed.
The required criteria for a good candidate were 
a high photon yield, in order to ensure a good ring reconstruction efficiency and accurate $\beta$ and charge measurements.  

The aerogel light yield depends on the tile thickness and on its optical
properties (refractive index and clarity).
The light yield has been evaluated from the analysis of helium samples
collected in 2003 and from the analysis of proton data samples gathered in
2002 with different beam momentum between 5 and 13\,GeV/c \cite{bib:pereiraLY}.
\begin{figure}[htb]
\begin{center}
\vspace{-0.5cm}
\begin{tabular}{cc}
\scalebox{0.22}{\includegraphics[bb=0 -20 595 567]{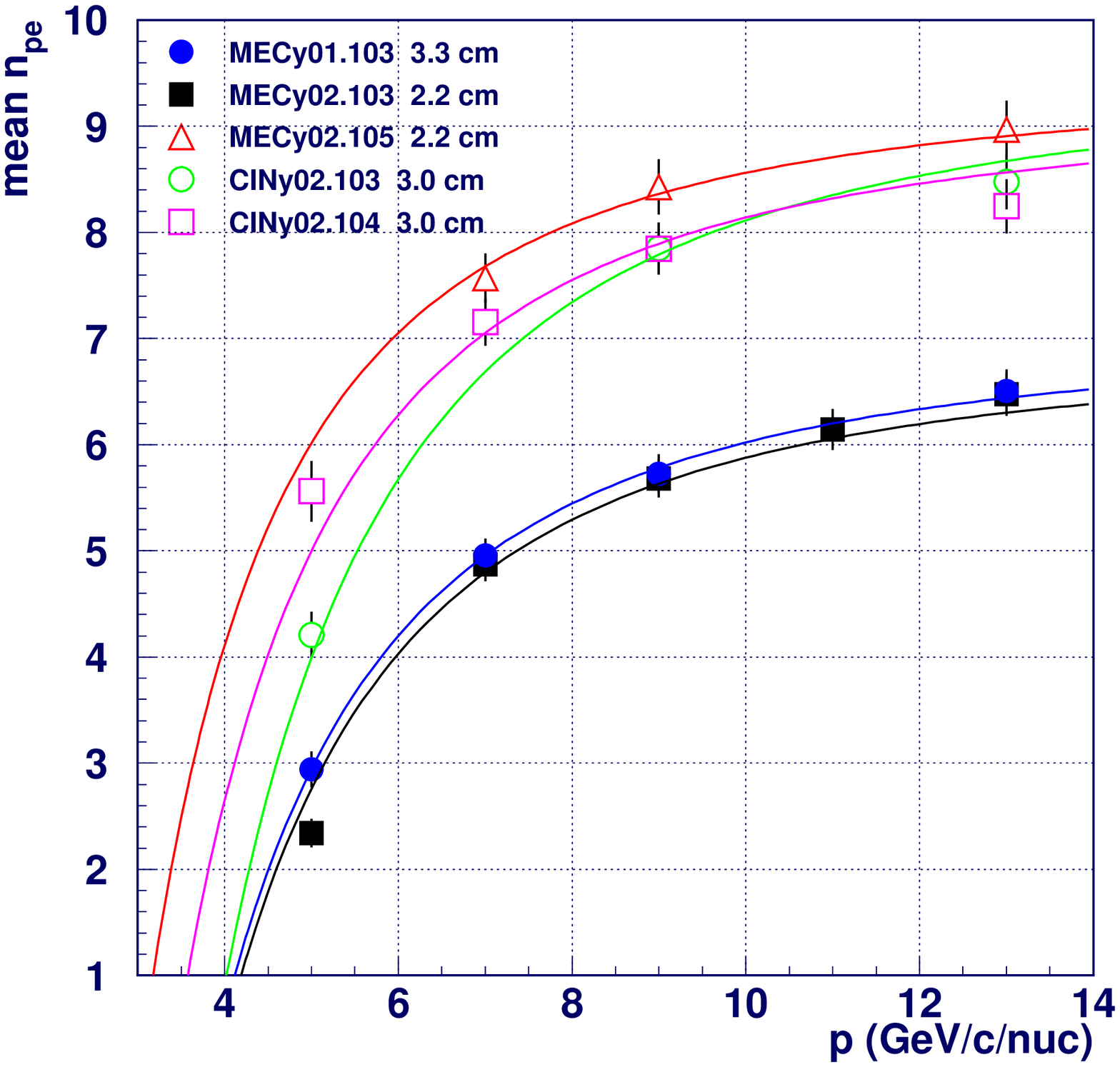}}
&
\hspace{-1.0cm}
\scalebox{0.3}{\includegraphics{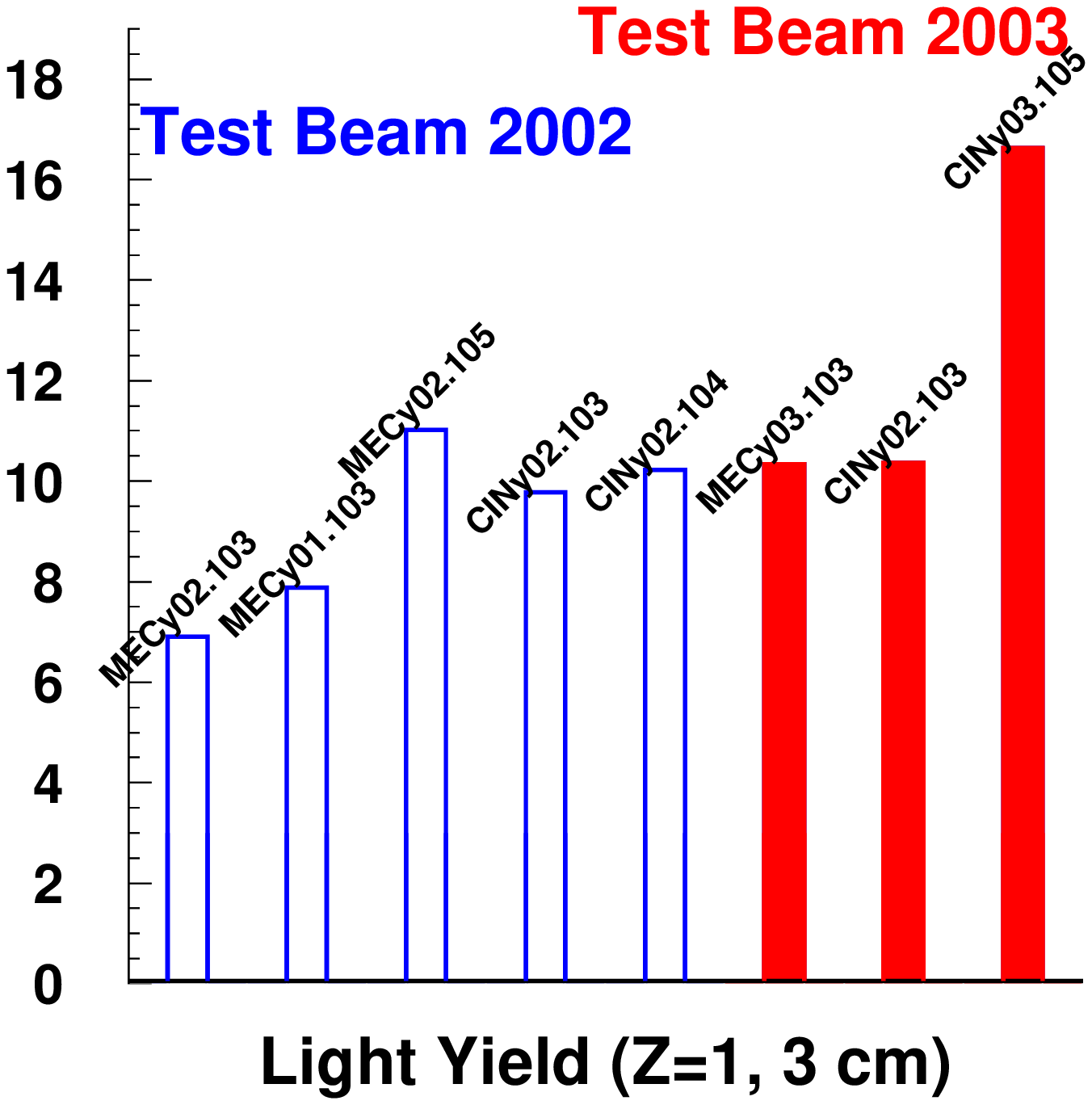}}
\end{tabular}  
\caption{Light yield as function of proton beam momentum for the different
  aerogel samples tested in 2002 (left). Light yield comparison based on test
  beam data 2002 and 2003. All values were extrapolated for fully cointained rings generated
  by a particle with $\beta$$\sim$1 and in an aerogel radiator with a common thickness of 3\,cm (right).} \label{fig:npe}
\vspace{-0.5cm}
\end{center}
\end{figure}

\mbox{Figure \ref{fig:npe}} (left) shows the evolution of the light
yield of the different aerogel samples tested in 2002 with the proton beam
momentum. A fit to each set of data was applied and the light yield for a
proton with $\beta$$\sim$1, generating fully contained rings in a radiator
with a common thickness of 3\,cm was extrapolated. Right plot of the same figure shows the normalized to 3\,cm
thickness light yield for the different aerogel samples tested in 2002 and
2003. Two interesting features are enhanced. In one hand, the same sample of
CINy02.103 was used in both years and its light yield analysis shows the same
value which proves the setup stability and the aerogel good performance after
one year period; on the other hand it is notorious the highest signal comes from a CIN sample produced in 2003 with $1.05$ refractive index reflecting the very
good clarity ($\sim0.0055~\mu$m$^4$/cm) of the aerogel batch.

\begin{figure}[htb]
\begin{center}
\vspace{-0.5cm}
\begin{tabular}{cc}
\scalebox{0.27}{%
\begin{overpic}{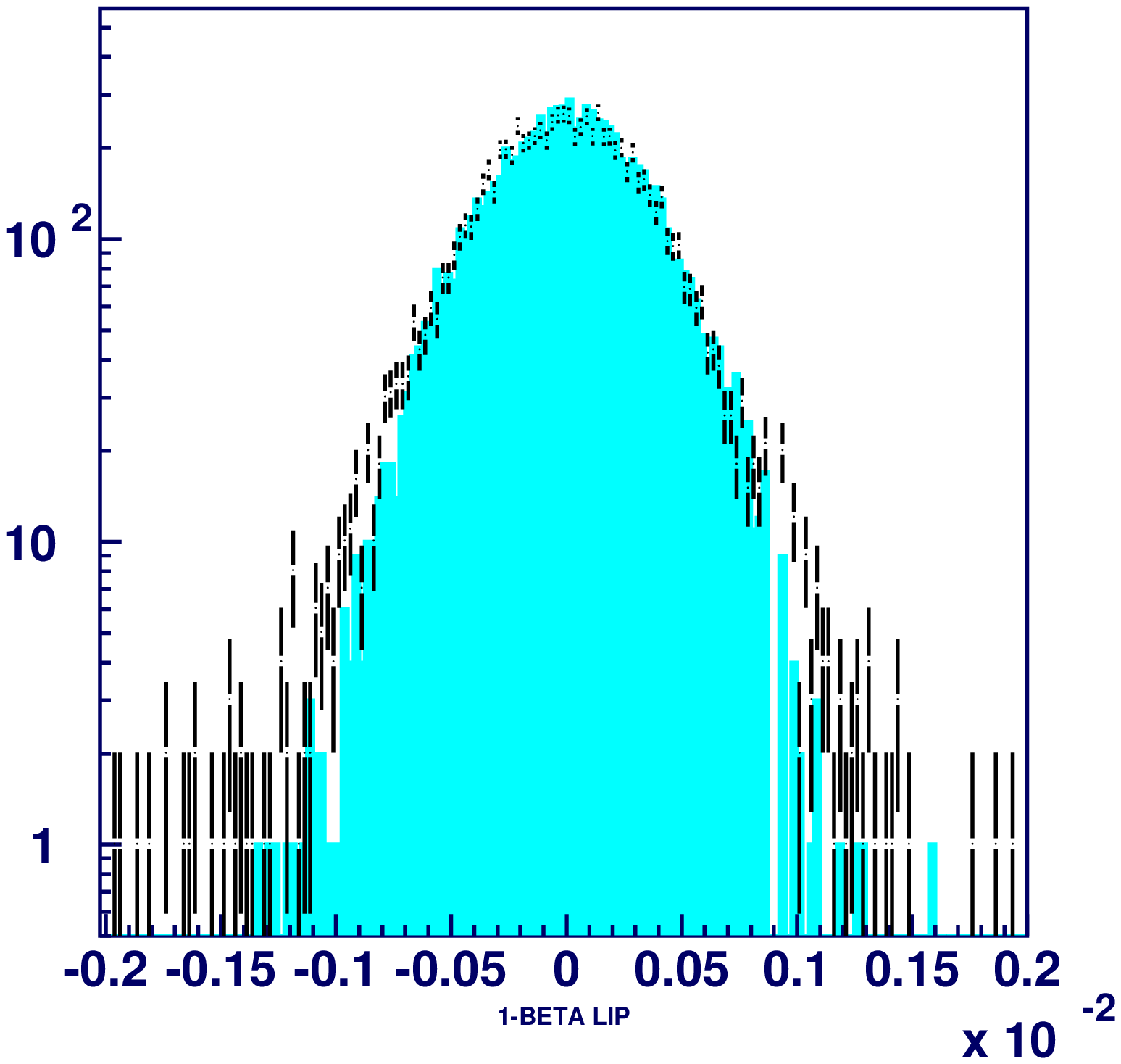} 
\put(67,75){\huge{$\bullet$}}
\put(71,75){\huge{DATA}}
\put(65,70){\myframebox{.03}{cyan}{~}}
\put(71.,70){\huge{MC}}
\end{overpic}
}
&
\hspace{-0.8cm}
\scalebox{0.27}{\includegraphics{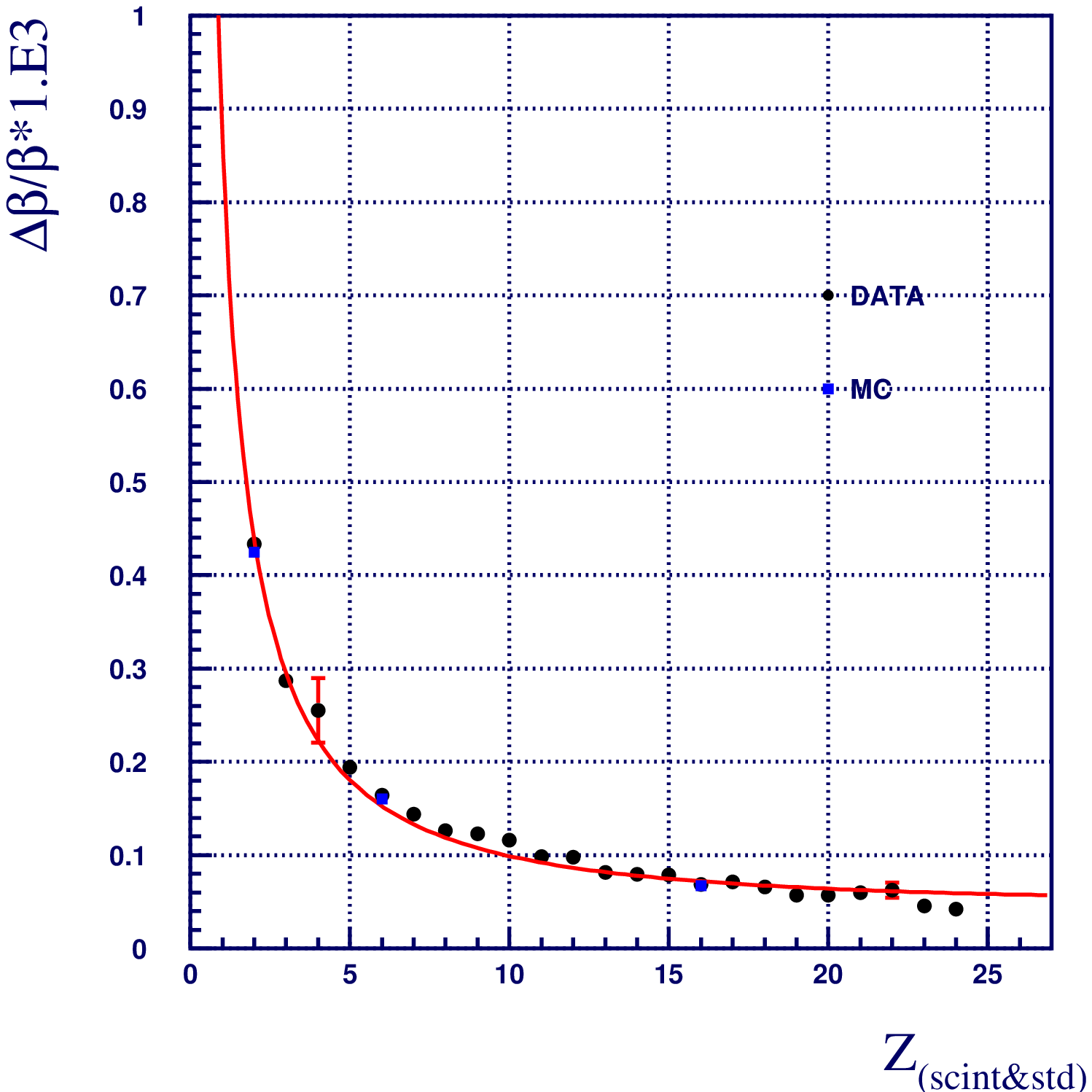}}
\end{tabular}  
\caption{Comparison of the $(\beta-1)*10^3$ distribution for helium data
  (black dots) and simulation (shaded) (left). Evolution of the $\beta$ resolution with the
  charge obtained for the same aerogel radiator. Simulated points for Z=2, 6,
  16 are marked with full squares (right). \label{fig:betatb03}}
\vspace{-0.5cm}
\end{center}
\end{figure}

The resolution of the $\beta$ measurement, obtained as explained in
\mbox{Section \ref{BETAZ}}, was estimated using a Gaussian fit
to the reconstructed $\beta$ spectrum, shown in left plot of \mbox{Figure
  \ref{fig:betatb03}} for helium nuclei. Data were collected with the aerogel
radiator CINy03.105, 2.5\,cm thick together with an expansion height of
35.31\,cm. The events shown correspond to particles inciding vertically and
generating fully contained rings. The beta reconstructed from a simulated helium data
is also shown (shaded histogram) superimposed with good agreement between data and Monte Carlo. 

The charge dependence of the velocity relative resolution for the same radiator is shown
in the right plot of \mbox{Figure \ref{fig:betatb03}}. The different charges were selected using
external and independent measurements performed by the silicon tracker
prototype and by the two scintillators. The observed resolution varies
according to a law $\propto$1/Z, as it is expected from the charge dependence
of the photon yield in the \CK\ emission, up to a saturation limit set by the
pixel size of the detection unit cell. The function used to perform the fit
is the following:
\begin{equation}
\sigma\left(\beta\right)=\sqrt{\left(\frac{A}{Z}\right)^2+B^2}
\end{equation}
where, $A$ means the $\beta$ resolution for a singly charged particle while $B$
means the resolution for a very high charge generating a large number
of hits, here the resolution is dominated by the pixel size as mentioned before.
The fitted values are $A=0.872\pm0.003$ and $B=0.047\pm0.001$ for the run
conditions stated above. 
Simulated data points for Z=2, 6, 16 are marked upon the same plot with full
squares. Once more the agreement between data and Monte Carlo measurements
for charges different of Z=2 is good.

The beta resolution for a $\beta\sim$1, helium nuclei impacting in each of
the aerogel samples tested in 2003 are summarized in \mbox{Table
  \ref{betares}}. The results are for a common expansion height extrapolated
from the values measured at the adjusted heights.

All the tested radiators fulfill the RICH requirement for $\beta$ measurement.

\begin{figure}[htb]
\begin{center}
\vspace{-0.5cm}
\begin{tabular}{cc}
\scalebox{0.24}{%
\includegraphics{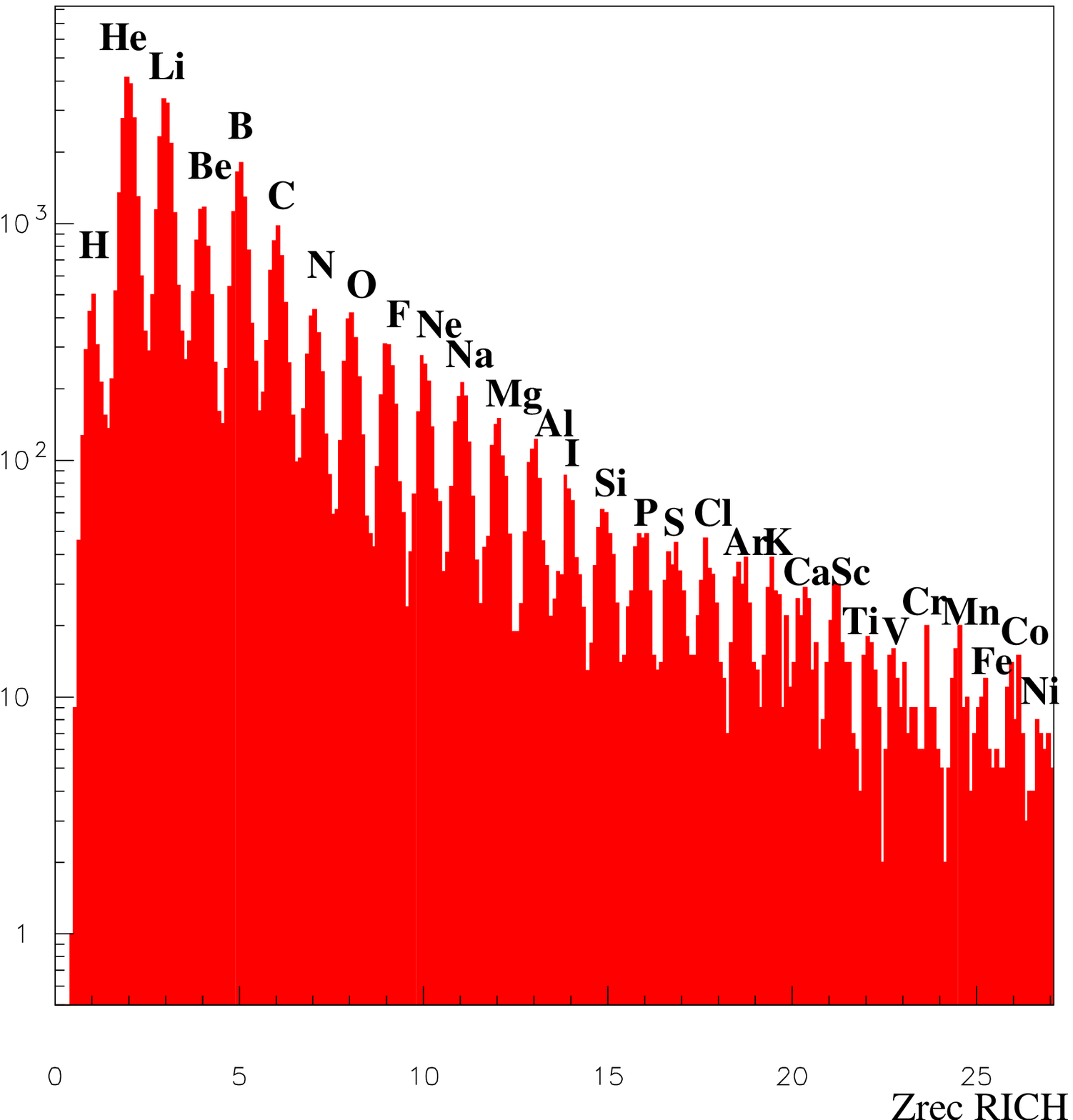} 
}
&
\hspace{-0.5cm}
\scalebox{0.24}{\includegraphics[bb=0 0 567 567]{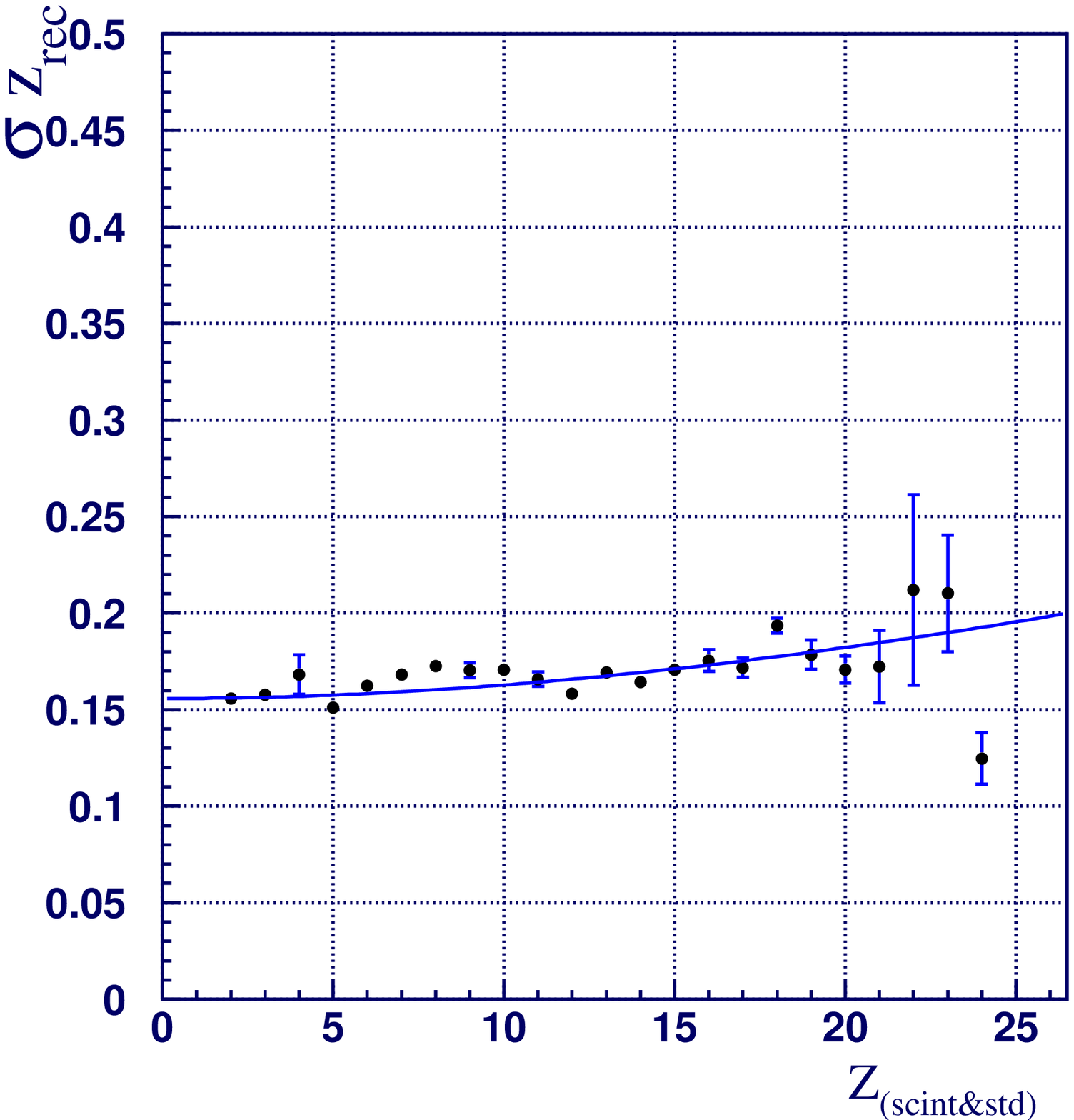}}
\end{tabular}  
\caption{Charge peaks distribution measured with the  RICH prototype using
  a n=1.05 aerogel radiator, 2.5\,cm thick. Individual peaks are identified
  up to Z$\sim$26 (left). Charge resolution versus particle Z for the same
  aerogel radiator. The curve gives the expected value estimated as explained
  in the text (right). \label{fig:chgtb03}}
\end{center}
\end{figure}

\begin{table}[htb]
\begin{center} 
\begin{tabular}{|c|c|c|c|}
\hline
\multicolumn{4}{|c|}{$\beta$ resolution for Z=2, H=33.5\,cm}\\
\hline
radiator & CIN103 & MEC103 & CIN105\\
\hline
$\sigma(\beta)\times10^3$ & 0.421$\pm$0.003 & 0.435$\pm$0.002 &
0.459$\pm$0.004 \\
\hline
\end{tabular}
\caption{Beta resolution for a helium particle with $\beta\sim$1 obtained for
  all the aerogel samples tested in 2003 and extrapolated for a common expansion
  height of 33.5\,cm.}
\label{betares}
\end{center}
\vspace{-0.6cm}
\end{table}
The distribution of the reconstructed charges in an aerogel radiator of n=1.05,
2.5\,cm thick is shown in left plot of \mbox{Figure \ref{fig:chgtb03}}. The
reconstruction method used was the one described in \mbox{Section
  \ref{BETAZ}}. The spectrum enhances a structure of well separated
individual charge peaks over the whole range up to iron (Z=26). This spectrum
was measured for a beam selection of A/Z=9/4.

The charge resolution for each nuclei, shown in right panel of \mbox{Figure
\ref{fig:chgtb03}}, was evaluated through individual
Gaussian fits to the reconstructed charge peaks selected by the independent
measurements performed by the scintillators and silicon tracker detectors.
A charge resolution for proton events slightly better than $0.17$ charge
units is achieved and as expected the best charge resolution is provided by
this radiator due to its higher photon yield.

The charge resolution as function of the charge Z of the particle follows a
curve that corresponds to the error propagation on Z which can be expressed
as:
\begin{equation}
\sigma(Z)=\frac{1}{2}\sqrt{\frac{1+\sigma_{pe}^2}{N_0}+Z^2\left(\frac{\Delta N}{N}\right)_{syst}^2}.
\end{equation}
This expression describes the two distinct types of uncertainties that affect Z
measurement: the statistical and the systematic. The statistical term is independent of the
nuclei charge and depends essentially on the amount of \CK\ signal
detected for singly charged particles ($N_{0} \sim 14.7$) and on the
resolution of the single photoelectron peak ($\sigma_{pe}$). The systematic
uncertainty scales with Z, dominates for higher charges and is around 1\%. It
appears due to non-uniformities at the radiator level coming from variations
in the refractive index, tile thickness or clarity or due to non-uniformities
at the photon detection efficiency like PMT temperature effects or light guide
non-uniformities. 

The RICH goal of a good charge separation in a wide range of nuclei charges 
implies a good mapping and monitoring of the potential non-uniformities
present on the detector. In order to keep the systematic uncertainties
below 1\%, the aerogel tile thickness, the refractive index and the
clarity should not have a spread greater than 0.25\,mm, $10^{-4}$ and $5\%$, respectively; at the detection level a precise knowledge ($<\,5\%$ level) of the single unit cell photo-detection efficiency and gains is required. 

Runs with a mirror prototype were also performed and its reflectivity was
derived from data analysis. The obtained value is in good agreement with the
design value.

\section{Conclusions}
AMS-02 will be equipped with a proximity focusing RICH detector based
on a mixed radiator of aerogel and sodium fluoride, enabling velocity measurements with a resolution of about 0.1\% and extending the charge measurements up to the iron element.
Velocity reconstruction is made with a likelihood method. Charge reconstruction is made in an event-by-event basis. 
Evaluation of both algorithms on real data taken with in-beam tests
at CERN, in October 2003 was done.
The detector design was validated and a refractive index 1.05 aerogel was
chosen for the radiator, fulfilling both the demand for a large light yield
and a good velocity resolution. The RICH detetector is being constructed and
its assembling to the AMS complete setup is foreseen for 2008.

\end{document}